\documentclass[twocolumn,preprintnumbers,amsmath,amssymb,superscriptaddress]{revtex4-2}
\usepackage{eqnarray,amsmath,lipsum}
\usepackage{graphicx}% Include figure files
\usepackage{dcolumn}
\usepackage{mathtools}
\usepackage{soul,xcolor}
\setstcolor{red}
\usepackage{natbib}

\usepackage{color}
\usepackage{mathrsfs}
\definecolor{red}{rgb}{1,0,0}

\definecolor{blue}{rgb}{0,0,1}

\definecolor{green}{rgb}{0,1,0}

\setcitestyle{square}

\begin{document}
\preprint{APS}

\author {Samuel M. Soares}
\affiliation{São Paulo State University (Unesp), IGCE - Physics Department, Rio Claro - SP, Brazil}
\author{Lucas Squillante}
\affiliation{São Paulo State University (Unesp), IGCE - Physics Department, Rio Claro - SP, Brazil}
\author{Henrique S. Lima}
\affiliation{Centro Brasileiro de Pesquisas Físicas, Rua Xavier Sigaud 150, Rio de Janeiro-RJ 22290-180, Brazil}
\author{Constantino Tsallis}
\affiliation{Centro Brasileiro de Pesquisas Físicas and National Institute of Science and Technology for Complex Systems, Rua Xavier Sigaud 150, Rio de Janeiro 22290-180, Brazil}
\affiliation{Santa Fe Institute, 1399 Hyde Park Road, Santa Fe, NM 87501, USA}
\affiliation{Complexity Science Hub Vienna, Josefstädter Strasse 39, 1080 Vienna, Austria}
\author{Mariano de Souza}
\email{mariano.souza@unesp.br}
\affiliation{São Paulo State University (Unesp), IGCE - Physics Department, Rio Claro - SP, Brazil}

\title{Universally non-diverging Gr\"uneisen parameter at critical points}

\begin{abstract}
According to Boltzmann-Gibbs (BG) statistical mechanics, the thermodynamic response, such as the isothermal susceptibility, at critical points (CPs) presents a divergent-like behavior. An appropriate parameter to probe both classical and quantum CPs is the so-called Grüneisen ratio $\Gamma$. Motivated by the results reported in Phys.\,Rev.\,B \textbf{108}, L140403 (2023), we extend the quantum version of $\Gamma$ to the non-additive $q$-entropy $S_q$. Our findings indicate that using $S_q$ at the unique value of $q$ restoring the extensivity of the entropy, $\Gamma$ is universally non-diverging at CPs. We unprecedentedly introduce $\Gamma$ in terms of $S_q$, being BG recovered for $q \rightarrow 1$. We thus solve a long-standing problem related to the \emph{illusory} diverging susceptibilities at CPs.
\end{abstract}

\maketitle

\begin{figure}[!t]
\centering
\includegraphics[clip,width=0.75\columnwidth]{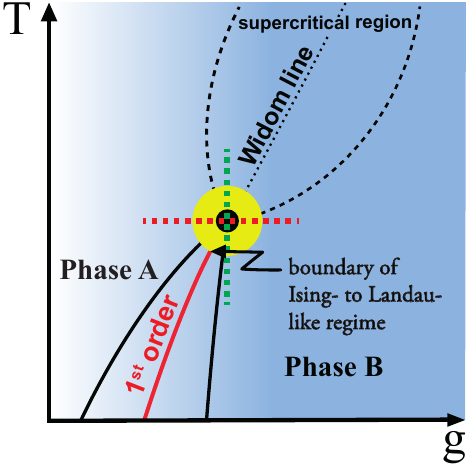}
\caption{\footnotesize Schematic generalized temperature $T$ \emph{versus} tuning parameter $g$ phase diagram showing that below the CP (black bullet) a phases coexistence region between phases A and B occurs, delimited by the spinodal lines (black solid line), which embodies a first-order transition line (red solid line). The yellow region around the CP indicates the Landau-like regime \cite{mdsprl2010,zacharias,mrb}. The red (horizontal) and green (vertical) dashed lines depict two possible ways of achieving the CP. The Widom line, embedded in the supercritical region, is also shown.} %variation of $g$ at $T_c$ and the variation of $T$ at $g_c$, respectively, in order to achieve the critical point at $(T_c,g_c)$.}} %The inset shows that upon varying $g$ at constant $T$ through the critical point, fluctuations of the order parameter take place, represented by the blue gradient.}
\label{Fig-1}
\end{figure}

\noindent\textbf{\emph{Introduction - }} The inherent phases competition in the immediate vicinity of both quantum and finite temperature $T$ critical points (CPs) gives rise to an entropy accumulation in that region of the phase diagram, cf.\,Fig.\,\ref{Fig-1}. Based on Boltzmann-Gibbs (BG) statistical mechanics, it is commonly discussed in the literature that in the critical regime the correlation length $\xi \rightarrow \infty$ and the thermodynamic response becomes singular \cite{Subir,stanley}. For the supercritical regime, $\xi$ is maximized upon crossing the Widom line \cite{ihme}, cf.\,Fig.\,\ref{Fig-1}. Such an intricate critical behavior of matter emerges in various distinct physical scenarios, including the Mott metal-to-insulator transition \cite{jap}, supercooled phase of water \cite{supercooled}, and quantum phase transitions \cite{unveiling}. It has been proposed that close to the finite $T$ Mott critical endpoint a crossover from Ising- to Landau-like regime (Fig.\,\ref{Fig-1}) takes place, most likely as a consequence of the intrinsic presence of defects and/or inclusions in real systems \cite{limelette}. From a historical perspective, based on Gibbs' seminal work, for diverging partition function, the coefficient of probability vanishes, so that \emph{the law of distribution becomes illusory} \cite{gibbs}. Over the last decades, the so-called Gr\"uneisen ratio, hereafter $\Gamma$, i.e., the ratio of the thermal expansion coefficient to the specific heat both at constant pressure, has been recognized as a useful \emph{tool} to explore critical manifestations of matter \cite{zhu}. Recently, some of us proposed a quantum version of $\Gamma$, namely $\Gamma^{0\textmd{K}}$, applicable at $T = 0$\,K, which can be considered as a \emph{compass} to quantify entanglement \cite{prbl}. In the latter, using the 1D Ising model under a transverse magnetic field (1DIMTF), we have illustrated that entanglement is enhanced near quantum CPs. We have shown that right at the quantum CP the Hellmann-Feynman theorem breaks down as a direct consequence of the high degeneracy in the critical regime. In this context, a key question emerges: is the thermodynamic response singular precisely at CPs in the frame of the non-additive $q$-entropy \cite{tsallistop}? The answer to this question constitutes one of the highlights of the present work, being this relevant since a milestone of the non-additive $q$-entropy refers to its applicability to long-range correlated systems \cite{tsallistop}.
We begin our analysis by recalling the generalized thermodynamical definition of $\Gamma$ \cite{mrb}:
\begin{equation}
\Gamma = \frac{\alpha_g}{c_g} = -\frac{\left(\frac{\partial^2 F}{\partial g\partial T}\right)}{T\left(\frac{\partial^2 F}{\partial T^2}\right)_g} = -\frac{1}{T}\frac{\left(\frac{\partial S_{BG}}{\partial g}\right)_T}{\left(\frac{\partial S_{BG}}{\partial T}\right)_g} = \frac{1}{T}\left(\frac{\partial T}{\partial g}\right)_{S_{BG}},
\label{gammaoriginal}
\end{equation}
where $\alpha_g$ is the thermal expansion coefficient, $c_g$ the specific heat, $g$ a tuning parameter (Fig.\,\ref{Fig-1}), $F$ the Helmholtz free energy, and $S_{BG}$ the BG entropy. For $T \rightarrow 0$\,K, it is clear that $\Gamma$ is undetermined so that a quantum version of $\Gamma$ is required, namely $\Gamma^{0\textmd{K}}$ \cite{prbl}:
\begin{equation}
\Gamma^{0\textmd{K}} = -\frac{\left(\frac{\partial^2 E_0}{\partial h\partial g}\right)}{h\left(\frac{\partial^2 E_0}{\partial h^2}\right)_g} = -\frac{\left(\frac{\partial S_N}{\partial g}\right)_h}{h\left(\frac{\partial S_N}{\partial h}\right)_g},
\label{prbl}
\end{equation}
where $E_0$ is the ground-state eigenenergy, $S_N$ the von Neumann entropy, and $h$ a distinct tuning parameter or energy scale associated with the quantum system of interest. It is to be noted that the definition of $\Gamma^{0\textmd{K}}$, cf.\,Eq.\,\ref{prbl}, can be adapted to suit the problem at hand, being thus reminiscent of the so-called $\beta'$ function $\beta'(g^*) = (L'/g^*)(dg^*/dL')$ in renormalization group theory, where $g^*$ is the observable and $L'$ the system's size \cite{blundell2}. Yet, based on Eqs.\,\ref{gammaoriginal} and \ref{prbl}, the definitions of both $\Gamma$ and $\Gamma^{\textmd{0K}}$ \emph{resemble} the so-called correlation function $\langle\langle M(x_2)M(x_1) \rangle\rangle \equiv \delta^2F[H]/\delta H(x_2)\delta H(x_1)...$, where $M$ is the magnetization at points $x_1$ and $x_2$ and $H$ the magnetic field intensity, discussed in the frame of Quantum Field Theory \cite{tsvelik}. Indeed, based on the fact that $\Gamma$ incorporates the entropy dependence in terms of two control parameters, cf.\,Eqs.\,\ref{gammaoriginal} and \ref{prbl}, it serves as a \emph{smoking-gun} to explore CPs. Yet, upon rewriting Eq.\,\ref{prbl} in terms of the generalized ratio $\lambda = g/h$, both numerator and denominator of $\Gamma^{0\textmd{K}}$ become equal, so that it suffices to analyze $dS_N/d\lambda$, cf.\,adopted in the present work. \vspace{0.10cm}

\noindent\textbf{\emph{Results, analysis, and discussions - }} To diverge or not to diverge? That is the question (to be answered)! We start recalling the $q$-generalized Boltzmann-Gibbs-von Neumann-Shannon entropy $S_q$ \cite{tsallistop}:
\begin{equation}
S_q(\hat{\rho}) = k\frac{1-\textmd{Tr}{\hat{\rho}}^q}{q-1},
\label{generalizedentropy}
\end{equation}
where $k$ is a positive constant, $q$ the entropic index, and $\hat{\rho}$ the density matrix operator. It is clear that for $q \rightarrow 1 \Rightarrow S_1 = S_N = -k\textmd{Tr}\hat{\rho}\ln{\hat{\rho}}$, which is the quantum version of $S_{BG}$ \cite{pretsallis}. It turns out that the Legendre transformations are $q$-invariant \cite{tsallisbook}, enabling us to generalize all thermodynamical relations straightforwardly in terms of $S_q$, so that $\Gamma$ (Eq.\,\ref{gammaoriginal}) can be expressed in terms of $S_q$ accordingly. Upon employing the $q$-generalized partition function $Z_q = \sum_{i = 1}^{W} [1 - (1-q)\beta E_i]^{1/(q-1)}$ \cite{tsallistop}, where $\beta$ is the inverse of the thermal energy and $E_i$ the $i^{\textmd{th}}$ eigenenergy, the $q$-dependent free energy is $F_q = -kT\ln_qZ_q$ \cite{tsallistop}, so that $\Gamma_q$ can be computed as:
\begin{equation}
\Gamma_q = -\frac{1}{T}\frac{\left(\frac{\partial^2 F_q}{\partial g \partial T}\right)}{\left(\frac{\partial^2 F_q}{\partial T^2}\right)_g},
\end{equation}
where $\ln_q{x} = (x^{1-q}-1)/(1-q)$ with $\ln_{1}{x} = \ln{x}$ \cite{tsallisbook}. Hence, it becomes evident that the $q$-dependence of $\Gamma_q$ lies on $E_i$. Yet, it is clear that for $q \rightarrow 1 \Rightarrow \Gamma_q = \Gamma$.

\emph{Canonical ensemble -} It is useful to analyze the case of a non-degenerate two-level system using $S_q$ in terms of the magnetic Gr\"uneisen parameter $\Gamma_{mag} = -1/T(\partial S/\partial B)_T/(\partial S/\partial T)_B$ \cite{modelsystems}, where $B$ is the magnetic field. We employ the $q$-dependent internal energy $U_q (\beta^*,\delta^*)$ reported in Ref.\,\cite{tsallistop}. We identify $\beta^*$ and $\delta^*$ as the thermal and magnetic energies, respectively. In analogy to $\Gamma_{mag}$, we compute the Gr\"uneisen parameter as $\Gamma^{two-level}_g \propto (\partial S_q/\partial \delta^*)_{\beta^*}/[\partial S_q/\partial (1/\beta^*)]_{\delta^*}$. Hence, using the results reported in Ref.\,\cite{tsallistop}, we obtain $\Gamma^{two-level}_g = 1/\delta^*$, which is in perfect agreement with $\Gamma_{mag} = 1/B$ reported by some of us for the Brillouin-like paramagnet \cite{modelsystems}.

\emph{Magnetic transition -} For $T = T_c$, $M$ is related to $H$ through the critical exponent $\delta$, i.e., $M \propto H^{1/\delta}$ \cite{stanley,tsallisbook}, being $\delta = (2q - 1)$ \cite{robledo}, enabling us to obtain $\Gamma_{mag}$ in terms of $q$, namely:
\begin{equation}
\Gamma_{mag,q} = -\frac{\alpha_H}{c_H} = \chi_T\frac{\alpha_M}{c_H} \propto \frac{1}{2q-1}H^{\{[1/(2q-1)]-1\}}\frac{\alpha_M}{c_H},
\label{ising}
\end{equation}
where $\alpha_H$ and $c_H$ are, respectively, the thermal expansion and the heat capacity, both at constant $H$, $\chi_T = (dM/dH)_T$ the isothermal magnetic susceptibility, and $\alpha_M = -\alpha_H/\chi_T$ the thermal expansion at constant $M$ \cite{stanley}. Note that for $q \rightarrow 1 \Rightarrow \Gamma_{mag,q \rightarrow 1} \propto \frac{\alpha_M}{c_H}$ $\propto$ $(\partial S/\partial M)_{T_c}/(\partial S/\partial {T_c})_H$, i.e., $\Gamma_{mag}$ for BG is recovered \cite{modelsystems}. Based on Eq.\,\ref{ising}, it becomes evident that $\Gamma_{mag} \propto \chi_T$, so that at CPs where $S_{BG}$ looses its thermodynamical extensivity, $\chi_T \rightarrow \infty \Rightarrow \Gamma_{mag} \rightarrow \infty$ in the frame of BG. As a matter of fact, a generalized Grüneisen parameter $\Gamma_g$ in terms of a generalized isothermal susceptibility $\chi_T^{\eta} = (d \eta/d h)_T$ can be proposed:
\begin{equation}
\Gamma_g = \frac{\chi_T^{\eta} \left(\frac{\partial S_{BG}}{\partial \eta}\right)_T}{c_h} = -\frac{1}{T}\frac{\left(\frac{\partial S_{BG}}{\partial h}\right)_T}{\left(\frac{\partial S_{BG}}{\partial T}\right)_{g}},
\end{equation}
where $\eta$ is the corresponding order parameter, such as $M$ or electric polarization $P$, and $h$ the conjugate quantity to the order parameter, e.g., magnetic or electric fields. Also, given that $\chi_T^{\eta} = -(\partial S_{BG}/\partial h)_T/(\partial S_{BG}/\partial\eta)_T$, a divergence of $\chi_T^{\eta}$ and, as a consequence, $\Gamma_g$, is expected at CPs in the frame of BG statistical mechanics. It is evident that any thermodynamic observable that depends on entropy diverges right at CPs using BG \cite{zhu,stanley}. In what follows, we demonstrate that upon considering the quantum version of $\Gamma_g$ in terms of $S_q$, such a divergence is suppressed for the unique value of $q$ that guarantees the thermodynamic extensivity of $S_q$: we may say that the theory is thus regularized.

\emph{The 1DIMTF -} For genuine quantum critical phenomena, $\Gamma^{0\text{K}}$ for the 1DIMTF can be computed generalizing $S_N$ in terms of $S_q$ in Eq.\,\ref{prbl}. In order to compute $\hat{\rho}$, it is worth recalling the 1DIMTF Hamiltonian \cite{pretsallis}:
\begin{equation}
\hat{H} = -\sum_{j = 1}^{N-1}(2\hat\sigma_j^x\hat{\sigma}_{j+1}^x + 2\lambda\hat{\sigma}_j^z),
\label{isinghamiltonian}
\end{equation}
where $\hat\sigma_j^x$ is the spin operator at the $j$-site along the $x$-axis, $\lambda = B/J$, $J$ the magnetic exchange coupling constant between nearest neighbors, and $\hat{\sigma}_j^z$ the spin operator along the $z$-axis.
\begin{figure}[!t]
\centering
\includegraphics[clip,width=0.92\columnwidth]{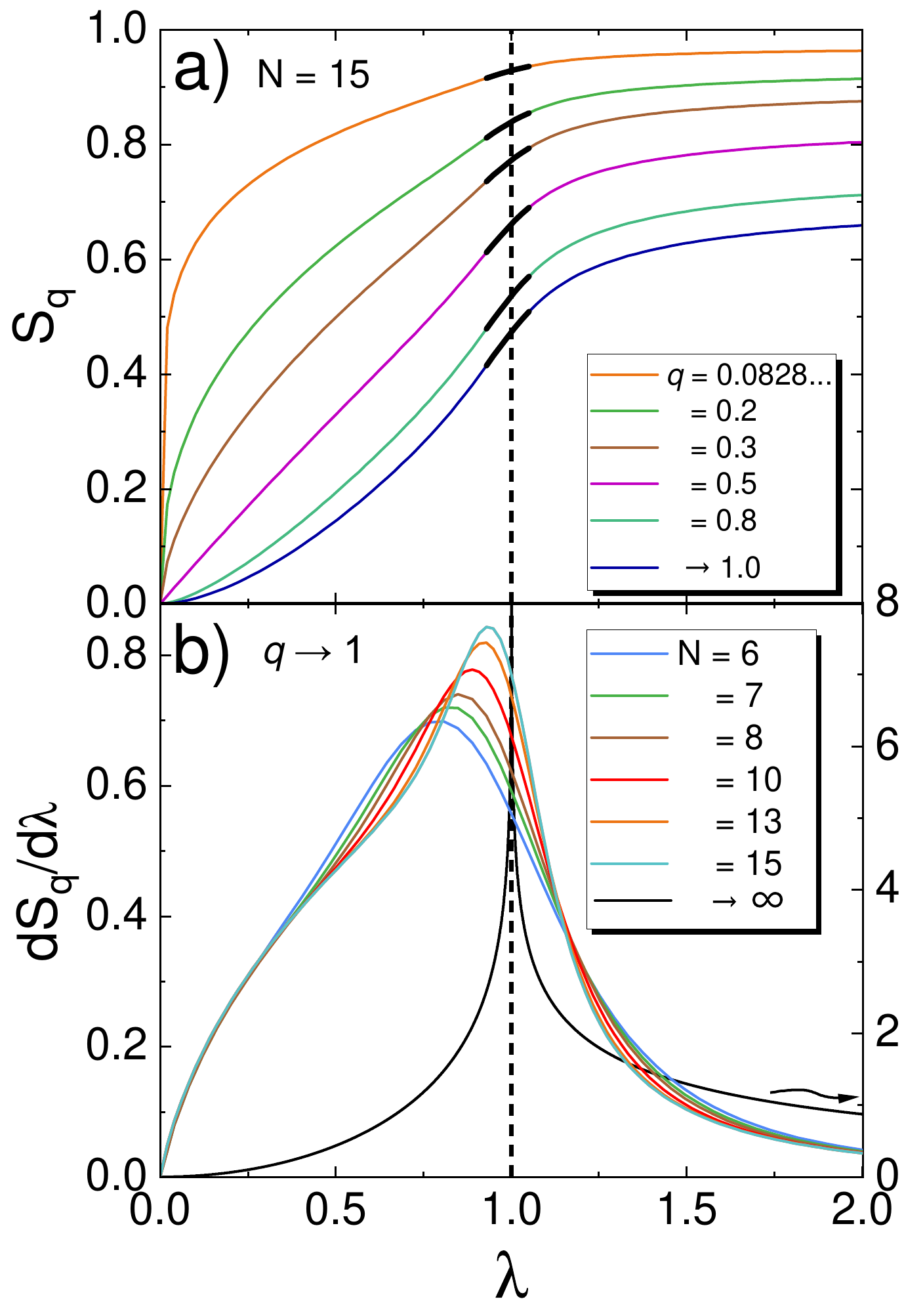}
\caption{\footnotesize a) $q$-generalized entropy $S_q$ for the 1DIMTF \emph{versus} $\lambda$ for $N = 15$ and various values of $q$, cf.\,label. The black solid lines highlight the derivative of $S_q$ in the vicinity of $\lambda = 1$. b) $dS_q/d\lambda$ \emph{versus} $\lambda$ for $q \rightarrow 1$ and various values of $N$, including our previous results regarding global entanglement for the thermodynamic limit \cite{prbl}.}
\label{Fig-2}
\end{figure}
By considering the 1DIMTF Hamiltonian (Eq.\,\ref{isinghamiltonian}), following the procedure discussed in Ref.\,\cite{neumann}, $\hat{\rho}$ can be computed, so that $S_q$ \emph{versus} $\lambda$ can be obtained for distinct values of $q$, cf.\,Eq.\,\ref{generalizedentropy}. Upon analyzing the behavior of $S_q$ \emph{versus} $\lambda$ in the vicinity of the 1DIMTF CP, i.e., for $\lambda = 1$, it becomes evident that the derivative of $S_q$ with respect to $\lambda$ is dramatically reduced upon decreasing the value of $q$, cf.\,black solid lines in Fig.\,\ref{Fig-2} a). Following discussions in Ref.\,\cite{pretsallis}, for $\lambda = 1$ the value $q = (\sqrt{9 + c^2} - 3)/c$, where $c$ is the central charge, ensures extensive $S_q$. For Ising-like models, $c = 1/2$, so that  $q = \sqrt{37} - 6 \simeq 0.0828\ldots$ .
\begin{figure}[!t]
\centering
\includegraphics[clip,width=0.92\columnwidth]{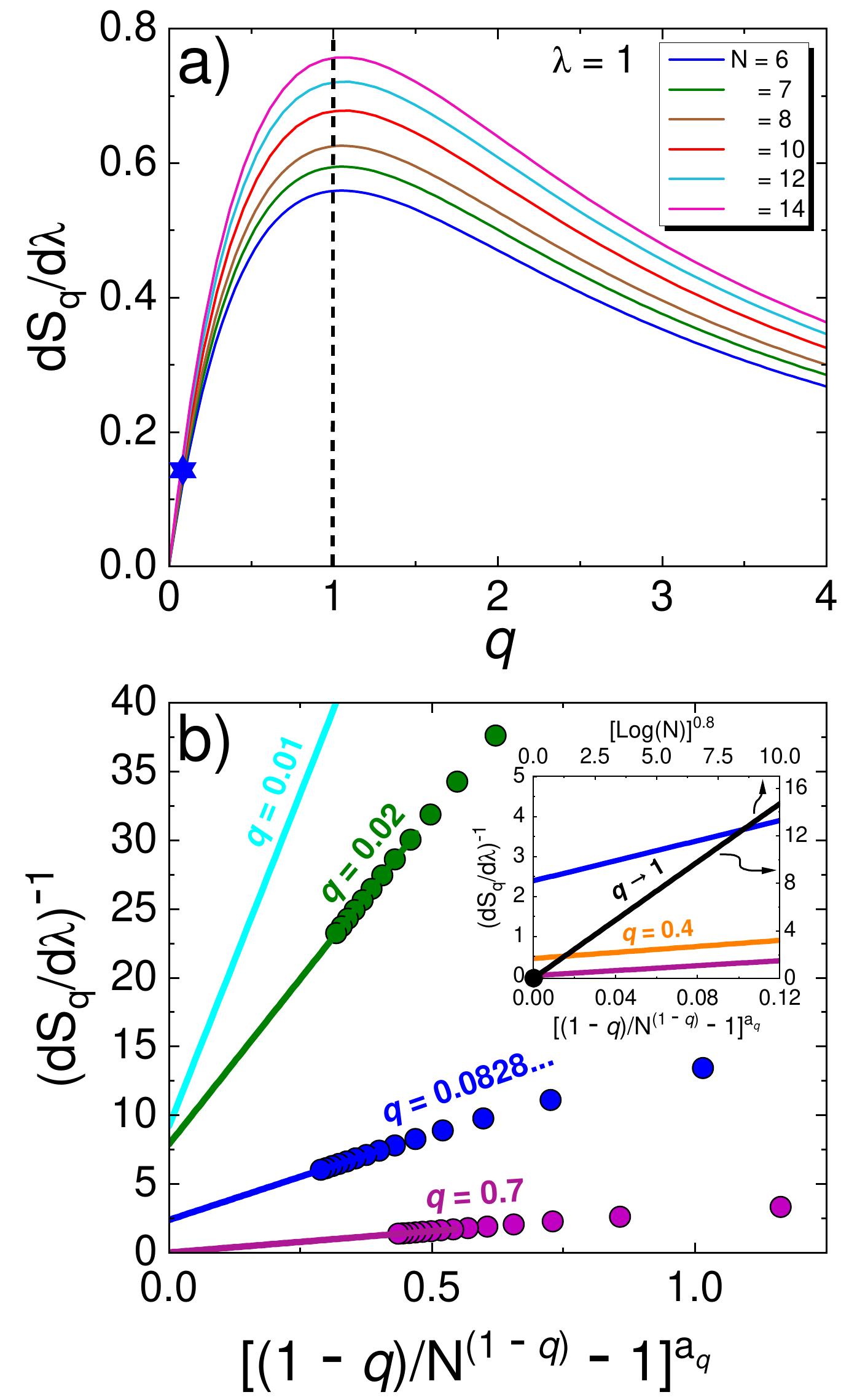}
\caption{\footnotesize a) $dS_q/d\lambda$ \emph{versus} $q$ for $\lambda = 1$ and distinct values of $N$. The blue star indicates $q = 0.0828\ldots$, which ensures extensivity of $S_q$ at the CP for the 1DIMTF \cite{pretsallis}. The vertical black dashed line indicates $q = 1$, which corresponds to the regime in which $dS_q/d\lambda$ is maximized. b) $(dS_q/d\lambda)^{-1}$ \emph{versus} $[(1-q)/N^{(1-q)}-1]^{a_q}$ with $N$ ranging from 2 to 15 for various values of $q$, cf.\,label. A linear fitting was employed in all data set for $N \geq$ 8 with $a_q =$ 0.37 (cyan), 0.44 (green), 0.50 (blue), and 0.58 (magenta). Inset: $(dS_q/d\lambda)^{-1}$ \emph{versus} $[(1-q)/N^{(1-q)}-1]^{a_q}$ (left and bottom axes) with $a_q =$ 0.55 (orange color line) and $(dS_q/d\lambda)^{-1}$ \emph{versus} [1/Log($N$)]$^{0.8}$ (right and top axes) for $q \rightarrow 1$ (black solid line). For $q = 0.4$, $(dS_q/d\lambda)^{-1}$ is finite because $N$ is finite, i.e., for $N \rightarrow \infty$ $\Rightarrow$ $(dS_q/d\lambda)^{-1} \rightarrow 0$. The black bullet at the origin indicates $dS_q/d\lambda \rightarrow \infty$ for $N \rightarrow \infty$ and $q \rightarrow 1$, cf.\,discussions in Ref.\,\cite{prbl}.}
\label{Fig-3}
\end{figure}
Note that we have considered the number of lattice sites $N = L \equiv$ block size in all calculations. Regarding Fig.\,\ref{Fig-2} b), upon increasing $N$, the maximum of $dS_q/d\lambda$ is shifted towards $\lambda = 1$, being $dS_q/d\lambda \rightarrow \infty$ for $N \rightarrow \infty$ (thermodynamic limit) and $q \rightarrow 1$ \cite{prbl}. Also, for small values of $N$, $dS_q/d\lambda$ is not maximized for $\lambda = 1$, cf.\,vertical dashed line. As $N$ is increased, the maximum of $dS_q/d\lambda$ approaches $\lambda = 1$. This is corroborated by the results depicted in Fig.\,\ref{Fig-3} a) of $dS_q/d\lambda$ \emph{versus} $q$ for $\lambda = 1$ and various values of $N$. For $N \rightarrow \infty$ and $q \rightarrow 1$, $S_q \rightarrow S_N$ \cite{fazio} and $dS_q/d\lambda \rightarrow \infty$, which is in perfect agreement with the results reported in Refs.\,\cite{prbl,asadian} considering global entanglement. Note that $dS_q/d\lambda$ is maximized for $\lambda = 1$ and $q \rightarrow 1$ [vertical dashed line in Fig.\,\ref{Fig-3} a)] due to the high number of accessible states. This is merely a consequence of the enhanced Hilbert space in this regime, which in turn is reflected on $\hat{\rho}$ \cite{hilbertbook}. Figure\,\ref{Fig-3} b) depicts $(dS_q/d\lambda)^{-1}$ \emph{versus} $[1/\ln_q{(N)}]^{a_q}$ for $q =$ 0.01, 0.02, 0.0828$\ldots$, and 0.7. The inset of Fig.\,\ref{Fig-3} b) shows $(dS_q/d\lambda)^{-1}$ \emph{versus} [1/Log($N$)]$^{0.8}$ demonstrating that for $N \rightarrow \infty$ and $q \rightarrow 1$, $(dS_q/d\lambda)^{-1} \rightarrow 0$, while for $q = 0.0828\ldots$ $(dS_q/d\lambda)^{-1}$ is finite for $N \rightarrow \infty$. The logarithmic-like divergence of $(dS_q/d\lambda)_{\lambda = 1}$ for $N \rightarrow \infty$ and particularly for $q \rightarrow 1$ asymptotically follows a linear behavior for large values of $N$ upon rewriting the abscissa as $1/[\ln_q{(N)}]^{a_q}$, cf.\,inset of Fig.\,\ref{Fig-3} b), and agrees with previous results regarding global entanglement \cite{asadian}. Our findings can be interpreted by making use of the following analogy: consider an Euclidian plane without rugosity, so that if one asks what its volume or its length is, the answer is, respectively, 0 and $\infty$. Of course, the appropriate question should be what is the area of its surface, which is indeed finite \cite{tsallisbook}. The same holds true upon considering fractals, i.e., Hausdorff dimension $d_f = \log{(N^*)}/\log{(1/E)}$, where $N^*$ is the number of elements that compose the particular geometry and $E$ the scale factor. More specifically, the $d$-dimensional fractal measure $M_d = 0$ for $d > d_f$, $M_d \rightarrow \infty$ for $d < d_f$, and finite only for $d = d_f$ \cite{tsallisbook}. A remarkable example of a fractal-like set refers to the configurations of Ising spins at CPs \cite{kada,wilson}. It is worth mentioning that the 2D Ising model can be modeled in fractal structures, such as the Sierpi\'nski carpet \cite{hebert}. In our analysis, $q$ plays the role of $d$, i.e., upon increasing $N$, extensivity is guaranteed for the \emph{special} $q$ value analogously to the fractal dimension upon increasing $N^*$. Our findings depicted in Fig.\,\ref{Fig-3} b) strongly indicate that for values of $q < 0.0828\ldots$ $\Rightarrow$ $(dS_q/d\lambda)^{-1} \rightarrow \infty$ and for $q > 0.0828\ldots$ $\Rightarrow$ $(dS_q/d\lambda)^{-1} \rightarrow 0$, making the value of $q = 0.0828\ldots$ \emph{special} because it guarantees that the Legendre structure of Thermodynamics is fulfilled and thus $(dS_q/d\lambda)_{\lambda = 1}$ remains finite, cf.\,Fig.\,\ref{Fig-3} b). In other words, upon using $q_{special} = 0.0828\ldots$ for the 1DIMTF, $dS_q/d\lambda$ remains finite at the CP even for $N \rightarrow \infty$. Hence, it is clear that the rate in which $S_q$ varies for $\lambda = 1$ upon increasing $N$ depends dramatically on $q$.
This is one of the key results of this work, which answers the previous posed question. It is remarkable that at CPs in the frame of BG, the underlying origin of the divergent-like behavior of $\Gamma$ lies on the \emph{illusory} \cite{gibbs} character of BG in such a regime, so that employing $S_q$ and the corresponding value of $q_{special}$ to ensure extensivity, $\Gamma$ is finite at CPs and presents a characteristic value for each system/model. For the 2D Ising model in the absence of $B$, the critical temperature plays the role of the critical field, so that we expect similar results at the CP. Our analysis is extendable to a wide variety of systems. Yet, since $\Gamma$ quantifies caloric effects, our findings suggest that non-diverging-like caloric effects are expected at CPs, as expected for real systems.

\textbf{\emph{Quintessence, conclusions, and outlook -}} The divergence of the thermodynamic response at CPs has been under a long debate \cite{stanley}. Of course, in real life, \emph{no one has ever measured an infinite value for any of these response functions} \cite{stanley}. Our findings indicate that the underlying origin of such a divergent thermodynamic response lies on the non-validity of BG at CPs \cite{tsallisbook}. It turns out that BG entropy is extensive only when short-range correlations are present, but since $\xi \rightarrow \infty$ at CPs, the emergence of long-range correlations in the system makes BG no longer valid at the precise CP. The latter can be also understood in terms of the enhancement of the number of microstates at CPs, i.e., $Z \rightarrow \infty$ \cite{gibbs}. To solve this issue, $S_q$ comes into play considering the corresponding system-dependent $q_{special}$ \cite{tsallistop,pretsallis}. Our findings suggest that the well-known divergent-like behavior of $\Gamma$ at CPs should be revisited in terms of $S_q$ \cite{zhu}. Yet, we interpret the constant value of $dS_q/dN$ for $q = q_{special}$ \cite{pretsallis} in analogy to $\beta'(g^*)$ in the scale-invariance regime \cite{blundell2}. The present analysis is not to be confused with Wilson's renormalization group theory \cite{wilson}. The latter basically acts on the parameters that appear in the Hamiltonian, whereas the present analysis refers to quantities which vary in that space.

In summary, we have proposed a new form for the Grüneisen ratio to investigate CPs using the $q$-entropy. Our analysis for the non-degenerate two-level system restores $\Gamma_{mag}$ for the Brillouin-like paramagnet. For the 1DIMTF, our findings strongly indicate that the divergence of $\Gamma^{0\textmd{K}}$ at the CP is suppressed for $q = q_{special}$. Our results suggest that at CPs, $\Gamma^{0\textmd{K}}$ is finite and for $q \rightarrow 1$ its \emph{illusory} divergent-like behavior inherent to BG is restored. Our analysis of the Gr\"uneisen parameter in terms of $S_q$ demonstrates the relevance of the non-additive $q$-entropy to investigate CPs among other open problems. The present numerical results at the CP indicate that, for $N \rightarrow \infty$, $\Gamma^{0\textmd{K}}$ $\rightarrow$ $\infty$ for $q > q_{special}$, including $q \rightarrow 1$, and $\Gamma^{0\textmd{K}}$ $\rightarrow$ 0 for $q < q_{special}$, being $\Gamma^{0\textmd{K}}$ finite ($\simeq$ 0.4) only for $q = q_{special}$. By considering the 1DIMTF taking into account say first and second nearest neighbors, $q_{special}$ remains the same, being $\Gamma^{0\textmd{K}} (q = q_{special})$ finite though different from the present results for $\lambda = 1$ and $N \rightarrow \infty$. The same holds true upon considering spin values differing from 1/2. Our approach can also be extended to dynamical systems, which constitute part of ongoing projects.

\textbf{\emph{Acknowledgements - }}MdeS acknowledges partial financial support from the S\~ao Paulo Research Foundation – Fapesp (Grants 2011/22050-4, 2017/07845-7, and 2019/24696-0), National Council of Technological and Scientific Development – CNPq (Grants 303772/2023-9). MdeS acknowledges RE Lagos-Monaco for discussions. CT acknowledges partial financial support from CNPq and Faperj. We also acknowledge LNCC (Brazil) for allowing us to use the Santos Dumont (SDumont) supercomputer. LS acknowledges IGCE for the post-doc fellowship. SMS and LS contributed equally to this work.
\medskip
\bibliographystyle{unsrt}

\end{document}